# Metal-insulator transition by isovalent anion substitution in $Ga_{1-x}Mn_xAs$: Implications to ferromagnetism

P.R. Stone[1,2,*], K. Alberi[1,2], S.K.Z. Tardif[2,†], J.W. Beeman[2], K.M. Yu[2], W. Walukiewicz[2] and O.D. Dubon[1,2,**]

[1]*Department of Materials Science & Engineering, University of California, Berkeley, CA 94720*
[2]*Lawrence Berkeley National Laboratory, Berkeley, CA 94720*

ABSTRACT

We have investigated the effect of partial isovalent anion substitution in $Ga_{1-x}Mn_xAs$ on electrical transport and ferromagnetism. Substitution of only 2.4% of As by P induces a metal-insulator transition at a constant Mn doping of $x=0.046$ while the replacement of 0.4 % As with N results in the crossover from metal to insulator for $x=0.037$. This remarkable behavior is consistent with a scenario in which holes located within an impurity band are scattered by alloy disorder in the anion sublattice. The shorter mean free path of holes, which mediate ferromagnetism, reduces the Curie temperature $T_C$ from 113 K to 60 K (100 K to 65 K) upon the introduction of 3.1 % P (1% N) into the As sublattice.

______________________________________________________________________

Because of their potential as both injectors and filters for spin-polarized carriers, ferromagnetic semiconductors have been proposed for use in spin-based electronics, or *spintronics*. These novel materials are formed by the substitution of a relatively small fraction of host atoms with a magnetic species. An important development toward the realization of ferromagnetic semiconductor-based devices was the discovery of ferromagnetism at temperatures up to 110 K in $Ga_{0.95}Mn_{0.05}As$ grown by low-temperature molecular beam epitaxy (LT-MBE)[1-3]. Ferromagnetism in $Ga_{1-x}Mn_xAs$ arises from hole-mediated inter-Mn exchange, and $T_C$ has been shown to increase with increasing concentration of substitutional Mn ($Mn_{Ga}$), or *x*. Thus far, Curie temperatures ($T_C$s) as high as 173 K have been reported for films of this canonical system[4]. Despite these remarkable observations, further increase in $T_C$ has been hindered by challenges in materials synthesis—namely, increasing *x* while avoiding the formation of second phases.

Increasing the magnitude of *p-d* exchange by tailoring the composition of the host semiconductor is another proposed, though significantly less explored, route by which to raise $T_C$[5, 6]. Utilization of semiconductors with band edges energetically closer to the Mn 3*d* levels increases hybridization between the host anion *p* states and Mn *d* states. However, this necessarily places the Mn acceptor level deeper in the band gap thus increasing localization of ferromagnetism-mediating holes. It has been proposed that dilute alloying of GaAs with GaP may yield a host in which the itinerancy of the mediating holes is maintained while *p-d* exchange is enhanced due to the shorter average Mn-anion bond length [5]. Recent calculations have, in fact, predicted an enhancement of $T_C$ by a factor of 1.5 due to this effect [6].

In this Letter, we report the synthesis and magneto-electronic properties of $Ga_{1-x}Mn_xAs_{1-y}P_y$ and $Ga_{1-x}Mn_xAs_{1-y}N_y$ ferromagnetic semiconductors with *y*<0.03. Even in this dilute limit the incorporation of P and N into $Ga_{1-x}Mn_xAs$ results in a strong *decrease* in $T_C$ with increasing *y*. Both quaternary alloys display a metal-insulator transition as a function of *y* even as $Mn_{Ga}$ is held constant.

All samples were synthesized using the combination of ion implantation and pulsed laser melting (II-PLM) [7, 8]. Briefly, $Ga_{1-x}Mn_xAs$ was synthesized by Mn ion implantation into semi-insulating GaAs followed by irradiation with a single pulse from a KrF ($\lambda$=248 nm) excimer laser at a fluence of 0.3 J/cm$^2$. Quaternary alloys were synthesized by performing further implantation of P or N ions prior to pulsed-laser melting[9]. Films were etched for 20 minutes in concentrated HCl to remove surface oxide layers [10]. Mn and P concentrations and substitutional fractions were determined by the combination of secondary ion mass spectrometry (SIMS) and ion beam analysis[7, 11]. For all samples the fraction of Mn atoms substituting for Ga is between 80-84%, which is comparable to that observed for LT-MBE grown films[4]. Importantly, the II-PLM process results in no interstitial Mn ($Mn_I$); the remainder of the Mn atoms is incommensurate with the lattice. The active nitrogen content *y* of $Ga_{1-x}Mn_xAs_{1-y}N_y$ films was determined by photomodulated reflectance spectroscopy[12, 13]. Magnetization measurements were performed using a SQUID magnetometer in an applied field of 50 Oe. DC transport was measured in the van der Pauw geometry using pressed indium contacts, which were determined by I-V measurements to be Ohmic for all temperatures presented in this work.

Before examining the properties of the $Ga_{1-x}Mn_xAs_{1-y}P_y$ system, we first review the magnetic properties of the endpoint compounds. The dependence of $T_C$ on *x* for $Ga_{1-x}Mn_xAs$ and $Ga_{1-x}Mn_xP$ is shown in Figure 1[4, 14]. For both materials we observe a monotonic increase in $T_C$ with *x*. The remarkable similarity of the dependence of $T_C$ with *x* (albeit shifted) between the two alloys in conjunction with the identical Mn $L_{3,2}$ X-ray magnetic circular dichroism (XMCD) lineshapes of $Ga_{1-x}Mn_xP$ and $Ga_{1-x}Mn_xAs$[15] suggests that similar ferromagnetic exchange mechanisms are operative in $Ga_{1-x}Mn_xP$ and $Ga_{1-x}Mn_xAs$ despite differences in *p-d* hybridization, Mn acceptor binding energies and carrier localization[16]. Focusing on the $Ga_{1-x}Mn_xAs$ data we see that the data for both the II-PLM-formed and optimally prepared LT-MBE-formed materials follow the same linear trend demonstrating that the use of II-PLM for materials synthesis has no discernable effect on $T_C$. Indeed, the static magnetic and magnetotransport properties of II-PLM-grown films are the same as those of LT-MBE-grown films[17]. Thus, we expect the properties of $Ga_{1-x}Mn_xAs_{1-y}P_y$ and $Ga_{1-x}Mn_xAs_{1-y}N_y$ presented here to be intrinsic to the materials rather than to be associated with any peculiarity of the II-PLM process.

The temperature dependence of the sheet resistivity for a series of $Ga_{0.954}Mn_{0.046}As_{1-y}P_y$ samples is presented in Figure 2(a). The sample with no phosphorous exhibits metallic transport as expected for $Ga_{1-x}Mn_xAs$ films synthesized by both LT-MBE and II-PLM with a sufficiently high $Mn_{Ga}$ concentration. The sample with $y=0.016$ also shows metallic behavior although its sheet resistivity ($\rho_{sheet}$) is somewhat higher than the pure $Ga_{1-x}Mn_xAs$ film. As $y$ continues to increase, we observe a metal-insulator transition (MIT), which is qualitatively similar to the doping-induced MIT in $Ga_{1-x}Mn_xAs$[18]. Using these standard definitions for the metallic and insulating phases of $Ga_{1-x}Mn_xAs$ the critical value of $y$ ($y_{crit}$) must be between 0.016 and 0.024 for $x=0.046$. An MIT is also observed when $Ga_{1-x}Mn_xAs$ is alloyed with N as shown in Figure 2(b) with $y_{crit} < 0.004$. The stoichiometrically similar alloys $Ga_{0.954}Mn_{0.046}As_{0.984}P_{0.016}$ and $Ga_{0.954}Mn_{0.046}As_{0.986}N_{0.014}$ show vastly different transport behavior. While alloying of the As sublattice with 1.4% nitrogen is sufficient to induce the MIT, the holes in the parallel phosphorus-alloyed sample remain itinerant. Hence, substitution of N for As in $Ga_{1-x}Mn_xAs$ has a much stronger effect on electrical transport than P. We emphasize that in both alloys these MITs are driven by the incorporation of *isovalent* species while the $Mn_{Ga}$ doping concentration is held at a constant value. Additionally, the electrical behavior of our $Ga_{1-x}Mn_xAs_{1-y}N_y$ samples is similar to those reported for LT-MBE-grown films[19-21].

The occurrence of a metal-insulator transition in $Ga_{1-x}Mn_xAs_{1-y}P_y$ and $Ga_{1-x}Mn_xAs_{1-y}N_y$ at these dramatically low anion alloying levels can be understood within the context of alloy disorder scattering of holes in an impurity band. We propose that an impurity band arises due to a valence band anticrossing (VBAC) interaction between the Mn impurity states and host valence band states. We note that an anticrossing interaction between magnetic impurities and the host semiconductor has recently been used to account for the sign and magnitude of the exchange energy in $Ga_{1-x}Fe_xN$[22]. Alteration of the GaAs valence band edges by either P or N is treated according to the virtual crystal approximation (VCA). The VCA approximation is well justified in these dilute ternary alloys as deviation from the linear dependence of the valence band location on composition is negligible. The anticrossing interaction is treated according to the **k·p** formalism in which the standard 6 x 6 Kohn-Luttinger matrix expressing the VCA-corrected valence band structure of the $GaAs_{1-y}P_y$ (or $GaAs_{1-y}N_y$) host is augmented with the six wavefunctions of the localized Mn *p*-states. Further details of the VBAC model can be found elsewhere[23, 24].

Transport within an impurity band can be either metallic or insulating in nature depending on the relative magnitude of the impurity band width, W, and the lifetime broadening of the hole energies, δE; for W>δE the scattering of holes is such that metallic transport is possible, while for W<δE transport occurs primarily by hopping conduction. More explicitly one can write for the lifetime broadening

$$\delta E = \frac{\hbar e}{\mu m_{eff}} \tag{1}$$

where $m_{eff}$ is the hole effective mass and $\mu$ is the hole mobility. The mobility is assumed to be dominated by scattering from both ionized impurities as well as alloy disorder, which we treat using a standard formula applicable to extended impurity band states,

$$\mu_{AD} = \frac{h^3 e}{8\pi^2 m_{eff}^2 k_f |V_{AD}|^2 \Omega(1-y)y}. \qquad (2)$$

$\Omega$ is the unit cell volume, and $k_f$ is the Fermi wavevector. $V_{AD}$ is the alloy disorder potential- i.e. the matrix element of the potential difference between the actual potential for the sites that are occupied by either As or P (N) and the average, composition weighted (VCA) potential, which should be evaluated using the wave functions of the Mn band. For transport within an impurity band $V_{AD}$ is taken to be the offset of the impurity band edges of the appropriate $Ga_{1-x}Mn_x$-pnictide endpoint compounds[25], which is estimated from experimentally-determined valence band offsets and Mn acceptor level positions. Figure 3 shows the impurity band width along with $\delta E$ calculated for $Ga_{0.959}Mn_{0.046}As_{1-y}P_y$ as a function of $y$ for $V_{AD}$ = 0.21 eV. According to these calculations the MIT should occur at approximately $y$ = 0.018 in $Ga_{0.959}Mn_{0.046}As_{1-y}P_y$, which is in good agreement with experiment (*c.f.* Figure 2(a)). While the Mn impurity band width decreases slightly with $y$, the driving force behind the MIT is the strong increase of $\delta E$ due to alloy disorder scattering. Estimation of the critical value of $y$ might be improved by including the effects of state broadening by Mn, which would decrease the mobility and thus shift the calculated MIT to slightly lower $y$.

Unlike the case of $Ga_{1-x}Mn_xAs_{1-y}P_y$, the exact value for $V_{AD}$ in $Ga_{1-x}Mn_xAs_{1-y}N_y$ is unknown since valence band offsets between GaAs and zincblende (ZB) GaN are not known. Given that $V_{AD}$~0.7 eV for wurtzite (WZ) GaN and that the bandgap of ZB GaN is 0.23 eV smaller than that WZ GaN, $V_{AD}$ should be between 0.47 and 0.7 eV depending on the exact positions of the band offsets. For $V_{AD}$=0.7 eV, the MIT is calculated to occur for $y$=0.0025 as is shown in Figure 3. Choosing smaller values of $V_{AD}$ (i.e., introducing a valence band offset between ZB GaN and GaAs) shifts the calculated value of the MIT towards $y = 0.004$, which is in better agreement with experiment. Nevertheless, our simple model reproduces and well describes the experimentally observed trend in the MIT in $Ga_{1-x}Mn_xAs_{1-y}P_y$ and $Ga_{1-x}Mn_xAs_{1-y}N_y$, namely, that significantly less N than P is necessary to induce a metal-insulator transition. The agreement of this model with our experimental data lends further support to the picture of impurity band ferromagnetism in $Ga_{1-x}Mn_xAs$ even when $x$ is high as 4.5%[26-28].

The scattering of holes by alloy disorder has a profound effect on $T_C$. Figure 4(a) shows thermomagnetic profiles for selected $Ga_{0.959}Mn_{0.041}As_{1-y}P_y$ films. Measurements were performed with the 50 Oe field applied parallel to either the in-plane $[1\bar{1}0]$ direction or the out-of-plane [001] direction; the addition of P to $Ga_{1-x}Mn_xAs$ films on a GaAs substrate results in a tensile-strain-induced rotation of the easy axis from in-plane to out-of-plane[9]. The films with $y$=0 and $y$=0.009 have $[1\bar{1}0]$ easy axes while the easy axis for films with $y \geq 0.024$ is perpendicular to the film. $T_C$ was determined by extrapolation of the steepest portion of the thermomagnetic curve corresponding to a sample's easy axis to zero magnetization resulting in an uncertainty of 2-3 K. Increasing the P concentration of the film causes a clear decrease in $T_C$ for the entire series of samples as shown in Figure 4(b). This trend is in agreement with theoretical calculations which show the importance of maximizing the hole mean free path ($l=\hbar k_f \mu/e$) to achieve the highest possible $T_C$ for a given $x$[29]. Isovalent anion substitution results in a decrease in $l$ since $\mu$ decreases as holes are scattered by an increasingly disordered potential landscape, thus lowering $T_C$. Indeed, $T_C$ drops by nearly a factor of 2 from $Ga_{0.954}Mn_{0.046}As$ to $Ga_{0.954}Mn_{0.046}As_{0.969}P_{0.031}$—to a value below that observed in $Ga_{0.954}Mn_{0.046}P$—with the substitution of only 3.1% of As atoms with P. Furthermore, we note that substitution of 1.0% N of As atoms decreases $T_C$ from 100 K to 65 K in $Ga_{1-x}Mn_xAs_{1-y}N_y$ (not shown).

Alloy disorder scattering has a strong effect on the magnitude of the saturation moment as well. Prior to the onset of the MIT, alloying the As sublattice with P decreases $T_C$ without changing the saturation moment of $4.2\pm0.2$ $\mu_B/Mn_{Ga}$ as shown in Figure 4 (c). A transition from the metallic to insulating state results in a reduction of the saturation moment to ~3 $\mu_B/Mn_{Ga}$ for $y=0.031$ despite the fact that the concentration of substitutional Mn remains the same. The random distributions of P (substituting As) and $Mn_{Ga}$ lead to regions in the film where stronger hole scattering by alloy disorder locally decouples $Mn_{Ga}$ moments from the global ferromagnetic exchange. Hence, when $y>y_{crit}$ $T_C$ is depressed further due to a decrease in the concentration of ferromagnetically active $Mn_{Ga}$ moments. Collectively these findings indicate that improvement of $T_C$ in $Ga_{1-x}Mn_xAs$ through isovalent anion substitution is fundamentally limited by alloy disorder scattering.

This work was supported by the Director, Office of Science, Office of Basic Energy Sciences, Division of Materials Sciences and Engineering, of the U.S. Department of Energy under Contract No. DE-AC02-05CH11231. The authors thank T. Dietl for fruitful discussions. P.R.S. acknowledges support from NDSEG and NSF Fellowships.

REFERENCES

1      H. Ohno, Science **281**, 951 (1998).

2      H. Ohno, J. Magn. Magn. Mater. **200**, 110 (1999).

3      H. Ohno, *et al.*, Appl. Phys. Lett. **69**, 363 (1996).

4      T. Jungwirth, *et al.*, Phys. Rev. B **72**, 165204 (2005).

5      A. H. Macdonald, P. Schiffer, and N. Samarth, Nat. Mater. **4**, 195 (2005).

6      J. Masek, *et al.*, Phys. Rev. B **75**, 045202 (2007).

7      O. D. Dubon, *et al.*, Physica B **376**, 630 (2006).

8      M. A. Scarpulla, *et al.*, Appl. Phys. Lett. **82**, 123 (2003).

9      P. R. Stone, *et al.*, Physica B **401-402**, 454 (2007).

10     K. W. Edmonds, *et al.*, Appl. Phys. Lett. **84**, 4065 (2004).

11     C. Bihler, *et al.*, Phys. Rev. B **75**, 214419 (2007).

12     W. Shan, *et al.*, Phys. Rev. Lett. **82**, 1221 (1999).

13     K. M. Yu, *et al.*, J. Appl. Phys. **94**, 1043 (2003).

14     R. Farshchi, *et al.*, Solid State Commun. **140**, 443 (2006).

15     P. R. Stone, *et al.*, Appl. Phys. Lett. **89**, 012504 (2006).

16     M. A. Scarpulla, *et al.*, Phys. Rev. Lett. **95**, 207204 (2005).

17     M. A. Scarpulla, *et al.*, J. Appl. Phys. **103**, 073913 (2008).

18     F. Matsukura, *et al.*, Phys. Rev. B **57**, R2037 (1998).

19     R. Kling, *et al.*, Solid State Commun. **124**, 207 (2002).

20     G. Kobayashi, *et al.*, J. Cryst. Growth **301**, 642 (2007).

21     I. Oshiyama, T. Kondo, and H. Munekata, J. Appl. Phys. **98**, 093906 (2005).

22     W. Pacuski, *et al.*, Phys. Rev. Lett. **100**, 037204 (2008).

23　　K. Alberi, *et al.*, Phys. Rev. B **75**, 045203 (2007).

24　　K. Alberi, *et al.*, (unpublished).

25　　T. Dietl, in *Handbook on Semiconductors*, edited by S. Mahajan (Elsevier, Amsterdam, 1994), Vol. 3.

26　　K. S. Burch, *et al.*, Phys. Rev. Lett. **97**, 087208 (2006).

27　　J. Okabayashi, *et al.*, Phys. Rev. B **64**, 125304 (2001).

28　　L. P. Rokhinson, *et al.*, Phys. Rev. B **76**, 161201(R) (2007).

29　　S. Das Sarma, E. H. Hwang, and D. J. Priour, Phys. Rev. B **70**, 161203(R) (2004).

FIGURE CAPTIONS

**Figure 1** – Dependence of $T_C$ on $x$ for II-PLM grown $Ga_{1-x}Mn_xP$ and $Ga_{1-x}Mn_xAs$, as well as LT-MBE grown $Ga_{1-x}Mn_xAs$. The dashed black and grey lines are linear fits to the $Ga_{1-x}Mn_xAs$ and $Ga_{1-x}Mn_xP$ data respectively. LT-MBE $Ga_{1-x}Mn_xAs$ data from Ref. 4. Selected II-PLM $Ga_{1-x}Mn_xP$ data from Ref. 14.

**Figure 2** – (a) (main) $\rho_{sheet}$ as a function of temperature for $Ga_{0.954}Mn_{0.046}As_{1-y}P_y$. A magnification of the low resistivity range is shown in the inset to emphasize the lineshape of the metallic samples. (b) $\rho_{sheet}$ as a function of temperature for $Ga_{1-x}Mn_xAs_{1-y}N_y$. Thin films with $y=0$, $y=0.004$, and $y=0.010$ have $x=0.037$ while the film with $y=0.014$ has $x=0.046$ and should be compared to the $Ga_{1-x}Mn_xAs$ reference sample in panel (a) of this figure.

**Figure 3** – Mn impurity band width (grey line) and lifetime broadening of the hole energies for $Ga_{0.954}Mn_{0.046}As_{1-y}P_y$ (black dashed line) and $Ga_{0.954}Mn_{0.046}As_{1-y}N_y$ (black solid line) as a function of anion sublattice composition.

**Figure 4-** (a) Magnetization as a function of temperature for selected $Ga_{0.954}Mn_{0.046}As_{1-y}P_y$ films. Filled symbols correspond to data collected with the applied field parallel to the in-plane $[1\bar{1}0]$ direction. Open symbols show data collected with the applied field parallel to the [001] direction for films with out-of-plane easy axes, thus allowing for more accurate comparison of $T_C$. (b) Dependence of $T_C$ on $y$ for $Ga_{0.954}Mn_{0.046}As_{1-y}P_y$ for small $y$. The dashed grey line represents the $T_C$ of $Ga_{0.954}Mn_{0.046}P$ indicating where the data points must eventually converge for $y = 1$ and is extrapolated from Figure 1. (c) Dependence of the saturation magnetization as measured in a field of 50 kOe as a function of $y$. The grey shaded regions of panels (b) and (c) represent the range of $y$ in which the MIT occurs.

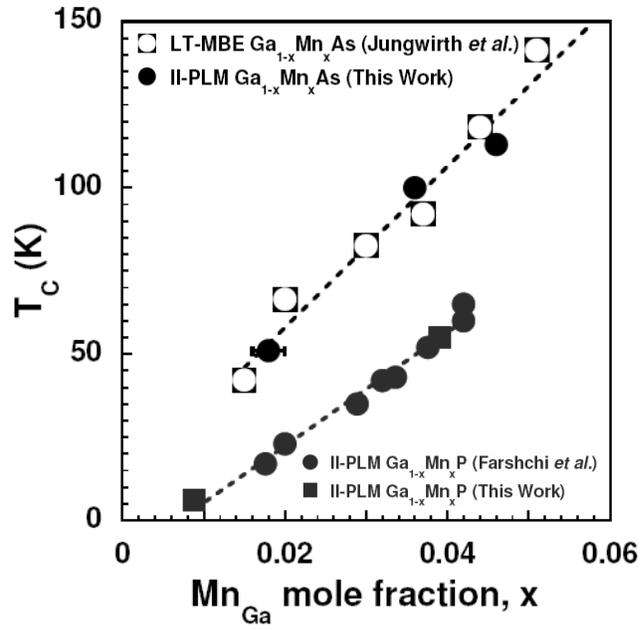

FIGURE 1- Stone *et al.*

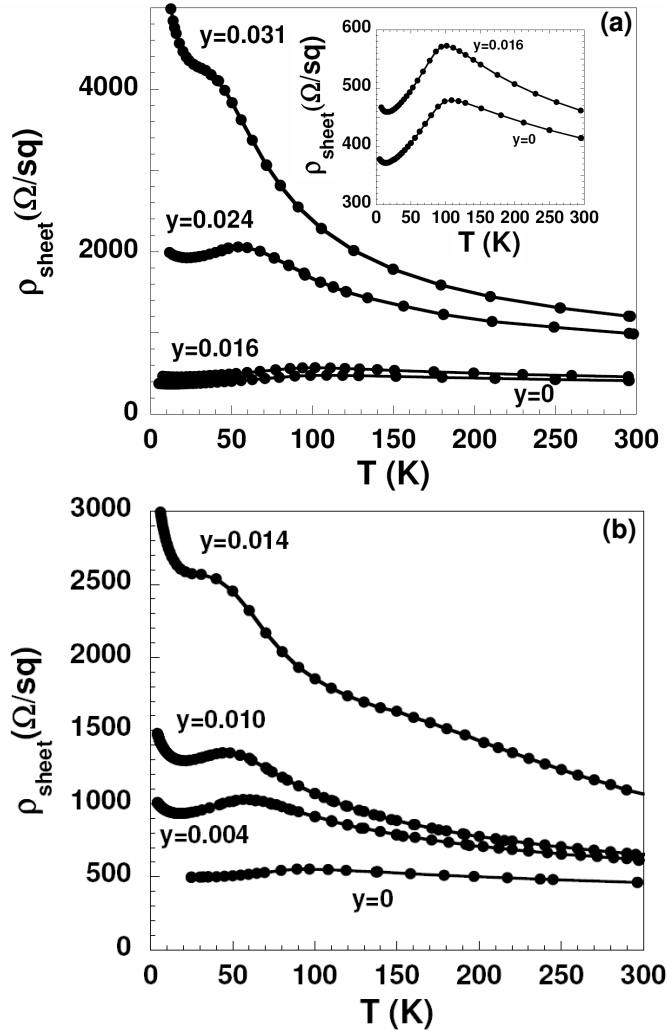

FIGURE 2- Stone *et al.*

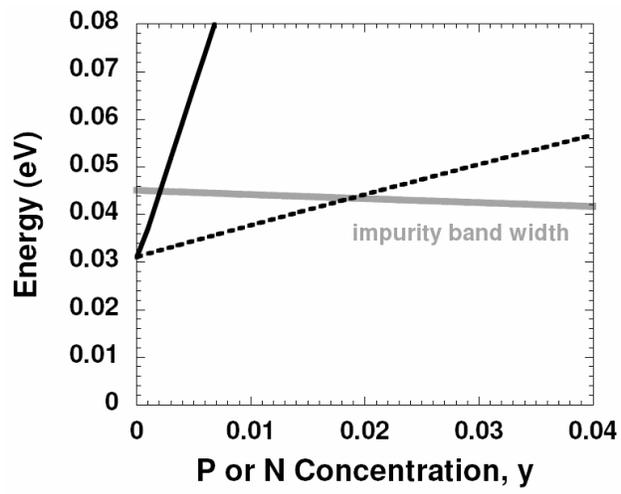

FIGURE 3 –Stone *et al*.

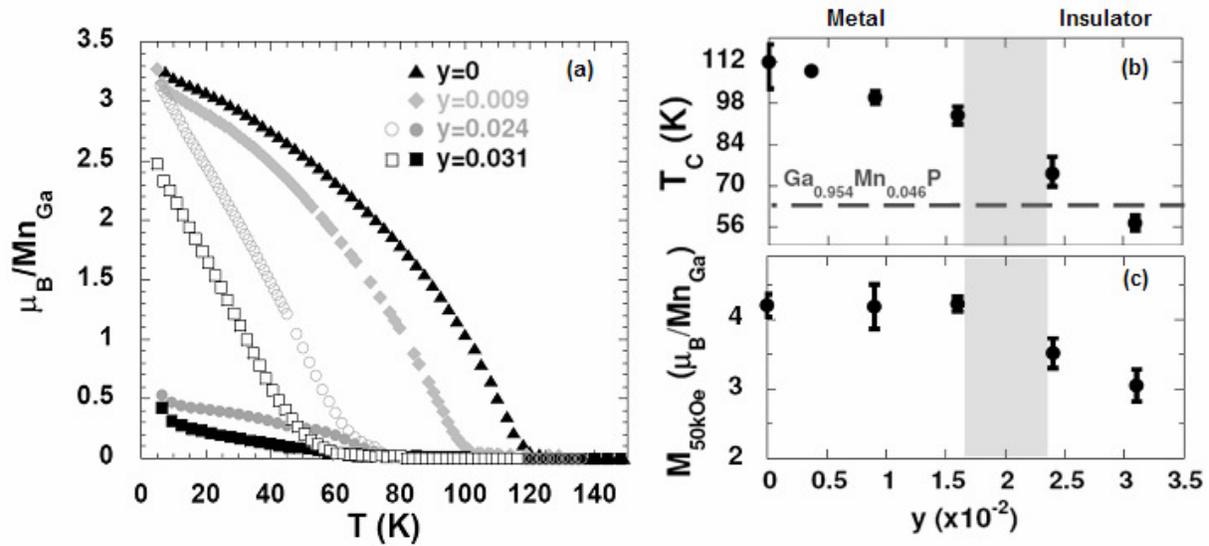

FIGURE 4 –Stone *et al*.